\documentclass{JHEP3}
\usepackage{amsmath}
\usepackage{enumerate}

\title {Notes on Spacetime Thermodynamics and the Observer-dependence of Entropy}

\author{ Donald Marolf\footnote{E-mail: {\tt marolf@physics.syr.edu}},\
Djordje Minic\footnote{E-mail: {\tt dminic@vt.edu}},\
and
Simon F. Ross\footnote{E-mail: {\tt S.F.Ross@durham.ac.uk}}\\
$^*$ Physics Department, UCSB, Santa Barbara, CA 93106.\\
$^\dagger$ Institute for Particle Physics and Astrophysics,
Department of Physics, Virginia Tech,
Blacksburg, VA 24061, U.S.A.\\
$^\ddagger$ Centre for Particle Theory, Department of Mathematical
Sciences, University of Durham, South Road, Durham DH1 3LE UK
}

\date{October, 2003}

\abstract{Due to the Unruh effect, accelerated and inertial observers differ in their description of a given quantum state.
The implications of this effect are explored for the entropy assigned by such observers to localized objects 
that may cross the associated Rindler horizon.  It is shown that the assigned entropies differ radically in the limit
where the number of internal states $n$ becomes large.   In particular, the entropy assigned by the accelerated
observer is a bounded function of $n$.  General arguments are given along with explicit calculations for free fields. 
The implications for discussions of the generalized second law and proposed entropy bounds are also discussed.}

\keywords{Unruh effect, Black Hole Entropy, Entropy Bounds} 
\preprint{hep-th/0310022}

\begin{document}

\section{Introduction}

The interaction of thermodynamics with gravitation has been a subject
of intense interest since the discovery of black hole entropy \cite{BHE} and
Hawking radiation \cite{haw}. The geometrical entropy associated with black hole
horizons is thought to provide an important clue to the quantum structure
of spacetime, and many attempts have been made to parse the corresponding riddle. The
approach most reminiscent of ordinary statistical mechanics attempts to identify a collection
of microstates whose counting reproduces this entropy. Despite some
successes, particularly in the context of string theory \cite{strbh}, we are still
far from having a concrete understanding of these states in the regime
where the classical geometrical description of a black hole is a good
approximation to the underlying physics (and we do not aim to improve
this situation in the present paper). 

Other authors have tried to
extend black hole entropy to
more general principles.  In particular, the conjecture that
the relation between entropy and area extends to any
event horizon 
(for example, to acceleration horizons in flat space)
has been recently championed by Jacobson~\cite{Jacob} and by
Jacobson and Parentani \cite{JP} who studied the associated first law of thermodynamics\footnote{Other work with this theme includes \cite{Pad} and references therein.}.  
An interesting output of this line of reasoning is the suggestion \cite{Jacob} that horizon entropy arises because gravity {\it
is} thermodynamics (or a special case thereof).   
Adopting an alternative approach, a number of
authors~\cite{Bek,erice,LS,tHooft} have argued that the consistency of
the second law of thermodynamics with a black hole entropy related to the area of the event
horizon yields a general principle restricting 
the allowed entropy of {\it any} system, be it black hole, matter, or other. 

Our aim in this paper is to explore a new relation between horizons
and entropy, which we believe is relevant to the discussion above. 
We will show that the entropy associated with a simple
localized matter
system in flat and otherwise empty space is not an invariant quantity defined by the
system alone, but rather depends on which observer we ask to measure
it. An inertial observer will assign the usual, na\"\i ve entropy given by the
logarithm of the number of internal states.  However, 
an accelerated observer (who sees the object immersed in a bath of
thermal radiation) will find the object to carry a {\it different} amount of entropy. 
Note that in the context we will consider both observers are able to describe the object
with the same degree of precision; the issue is {\it not} that our object is partially
hidden behind the Rindler horizon.

It is of course well known that the inertial and Rindler observers already ascribe a
different entropy to the Minkowski vacuum, as this is a thermal state
with divergent entropy \cite{entangle} from the Rindler
point of view. Considering both this fact and the background structures necessary for standard discussions of thermodynamics, 
Wald has argued for some time \cite{Wald1} (see also the last part of \cite{WaldR}) 
that entropy is an extremely subtle concept in general relativity -- 
even for ordinary matter systems -- and that we still lack the proper framework for a general discussion.
Our results are in complete agreement with this philosophy and may be considered a next small step in pursuit of this goal.
In particular, we now learn that the observer dependence of entropy is
far more than a simple shift of the zero point.   What is perhaps surprising from the point of view of
\cite{Wald1} is that the observers can disagree on the entropy of a localized object even when
they assign the same energy to all of its microstates.  

The fact that an accelerated
observer does not measure the usual na\"\i ve entropy for an inertial
matter system clearly has important consequences for thermodynamic discussions, 
and our motivation for investigating these questions is closely related to recent
explorations of such issues. 
Let us therefore recall the recent progress
\cite{MS1,MS2} in
clarifying issues related to the 2nd law as raised in
\cite{Bek,erice,LS,tHooft}. To review the basic question, suppose one
considers a black hole of temperature $T$ and mass $M$.  Classically, a mass $M$ black
hole will absorb any small object placed nearby\footnote{Though classically such a black hole will
have $T=0$.}.  If an object ($obj$)
carries energy $E_{obj}$ into the black hole the associated increase in black
hole entropy will be $dS=E_{obj}/T$ by the first law.  But what if the small object itself
has more entropy $S_{obj}$ than ${E_{obj}/T}$? Then such an absorption would
violate the generalized 2nd law.

But let us also recall that \cite{sorkin} argues that the generalized 2nd law will {\it in general}
be satisfied when the system outside the black hole can be described by quantum
field theory in curved spacetime.  Though issues of divergences and back-reaction prevent
this from being a rigorous proof, it is at least highly suggestive.
If correct, then there are two logical
alternatives. Either objects with $S_{obj} > E_{obj}/T$ cannot exist in any consistent
quantum field theory (in curved spacetime), or some quantum effect
must intervene to increase the entropy of the final
state. 

The latter conclusion was recently argued in \cite{MS1,MS2}. The
main point was that for $S> E/T$ either i) thermally produced objects
macroscopically indistinguishable from that which one is attempting to insert
into the given black hole will play an important role in the thermal
atmosphere of this black hole, or ii) the object of interest will be unable to freely
penetrate this black hole's thermal atmosphere.  There is a large overlap between
the two cases; for example, the presence of similar objects in the
thermal atmosphere can cause the new object one introduces to experience a
repulsive pressure. Such an effect might come either through an
explicit interaction or through statistical effects (e.g., the Pauli exclusion principle).
This repulsive pressure can prohibit our object from
falling into the black hole.

In the case where our object {\it can} freely penetrate the thermal
atmosphere, \cite{MS2} suggested how quantum effects
could still alter the final state.  The point is that, as mentioned above, 
the thermal atmosphere must contain a significant number
of such objects.  This is argued \cite{MS2} by computing the free energy
$F_{obj} =  E_{obj} - T S_{obj}$
of such an object at the Hawking temperature $T$ and finding that $F_{obj}$
is negative. Since the vacuum has zero free energy (i.e., a greater amount), 
at temperature $T$ our objects are more likely to
exist than not.

Suppose now that the black hole were to come to equilibrium (say, if placed
in a small reflecting cavity) with its thermal atmosphere.  Such an
equilibrium state would contain a significant flux of objects directed toward the 
black hole.  But since there is equilibrium, it must be that 
similar objects are also
radiated from the black hole at a significant rate.  Ref. \cite{MS2}
argues that this will be the case even for large $E_{obj}/T$ due to the fact
that the entropy $S_{obj}$ has been assumed even larger.

This effect appears to be sufficient to protect the second law when one drops a single 
$S_{obj} > E_{obj}/T$ object (or a small number of such objects) into a black hole in otherwise empty
space.  But there remains a final wrinkle to sort out: what if we measure
the flux of such objects being radiated from the black hole and then
arrange to beam our objects into the black hole at a higher rate?  Or,
suppose we place the black hole in a reflecting cavity, let it reach
equilibrium, and {\it then} send in another object with more than
$e^{E_{obj}/T}$ internal states?

Since the density of such objects in equilibrium may already be high,
this may require us to assume extremely weak interactions.  But this
is not a problem in principle, and in the present work we focus on the
case of free fields. In the limit of a large black hole, the problem
reduces to the study of a Rindler horizon in flat spacetime.  The
equilibrium state is just the Minkowski vacuum which, however, appears
thermally excited to uniformly accelerated (i.e., Rindler) observers.

Here, the observer-dependence of entropy becomes crucial. We find that
inertial and Rindler observers do {\it not} ascribe the same amount of
entropy to our object.  Inertial observers ascribe an entropy equal to
the logarithm of the number of internal states, as expected.  However, in the limit where the number of internal states $n$ is large,
a Rindler observer ascribes only an entropy $S_{acc} = E_{acc}/T$,\footnote{Here $E_{acc}$ (the energy measured by the {\it acc}elerated  observer) 
is the Killing energy associated with the
boost symmetry $\xi$ (i.e., the Rindler time translation).   The associated temperature $T$ is
given by $T = {\kappa \over 2 \pi} $ where
$\kappa$ is the surface gravity of $\xi$.  As usual, the normalization
of $\xi$ cancels so that $E_{acc}/T$ is independent of this choice.}\
suggesting consistency of the second law\footnote{Either the second
law for black holes in the limit of a large black hole, or the
``stationary comparison" second law for asymptotic Rindler
horizons \cite{JP}.} when such an object crosses the event horizon.
Thus, a consequence of our observer-dependence of entropy is
that allowing an object to fall across an event horizon {\it will}
plausibly respect
the generalised second law from the point of view of
the accelerated observer who remains outside of the horizon, no matter
how many internal states the object carries. 
However, our analysis stops short of being a proof for reasons related to the
unresolved issues with \cite{sorkin}; these are discussed further in
the concluding section.
(The inertial observer, on the other hand, never loses sight
of the object, so there can be no question of a violation of the
second law from their point of view.)

We will give a general argument for this
observer-dependence of the entropy in the next section. Our argument relies simply on
general points about thermodynamics and the relation of entropy to
statistical ensembles. We will see that the Rindler observer ascribes an entropy $S_{acc}
= E_{acc}/T$ to the object whenever it represents a small perturbation on the
thermal state, which in particular will be true in the limit of a
large number of microstates.  Section \ref{calc} then fleshes out
the detailed calculations in the case of free boson and free fermion
fields.  In particular, the relation between the Rindler energy $E_{acc}$ of
the object and the inertially measured energy $E_{inertial}$ is
calculated for a well-localized object.  With the proper normalization
of $E_{acc}$, and in the limit $E_{inertial} << T$, one finds $E_{acc} = E_{inertial}$ as one
would expect.  However, for larger $T$ the answer can be rather
different.  We close with some discussion in section \ref{disc}.

\section{Entropy and observers}

\label{gen}

The central point of this paper is to show that the entropy ascribed
to a localized physical system by different observers can differ, even when both
observers describe the region containing the system with the same resolution. 
We exhibit this in a particularly simple context,
considering the entropy of an object in flat space as seen by an
inertial or a Rindler observer. Note that we take the spacetime to be
exactly flat and explicitly ignore any possible gravitational effects from the objects we
discuss.  We will return to this point in section \ref{disc}.
In this section, we give a general
argument suggesting that the entropy measured by the Rindler observer will be
bounded by $E_{acc} /T$ and strongly arguing for this result in the limit of a large number of internal states.
Here $E_{acc}$ is the 
energy associated to the object by the Rindler observer and $T$ 
temperature seen by this observer.  In the following section, we give some more
detailed calculations for the case of single-particle excitations of
free bose or fermi fields.

{}From the inertial observer's point of view, the natural vacuum state
is the Minkowski vacuum  $|0_M\rangle$, and we describe the
excitation of this vacuum corresponding to the presence of some object
(in an undetermined microstate) by the density matrix
\begin{equation}
\rho_M = {1 \over n} \sum_{i=1}^n |1_M;i \rangle \langle1_M;i|,
\end{equation}
where the $|1_M;i\rangle$ denote Minkowski one-object states
describing the $n$ possible microstates of our object. The entropy
assigned to the object by the inertial observer is then as usual
\begin{equation}
S_M = - \mbox{Tr} \rho_M \ln \rho_M = \ln n.
\end{equation}

But 
what entropy will a Rindler observer assign
to this object? Note that the answer is {\it not} given by 
$-\mbox{Tr} \rho \ln \rho$ where $\rho$ is the 
density matrix corresponding to $\rho_M$ from the Rindler
observer's point of view.  This 
$-\mbox{Tr} \rho \ln \rho$ would include 
in addition a contribution from the
background of thermal acceleration radiation that the Rindler observer
may not wish to ascribe to the localized object.  One way to define the
entropy of the object from the Rindler observer's point of view is to
ask what entropy is carried out of the visible Rindler wedge when the
object crosses the Rindler horizon\footnote{The terms ``visible" and
``invisible" Rindler wedge will always be used in the context of what
is in causal contact with our chosen Rindler observer.  Of course, the
entire spacetime is visible to the inertial observer.}.  Thus, the
appropriate Rindler notion of entropy $S_{acc}$ {\it carried} by the object 
is the {\it difference} $\delta S$ between the
entropy in the visible wedge when the object of interest is present
and when it is not.  Similarly $E_{acc} = \delta E$ is the corresponding difference in
Killing energies.  We repeat that we work in the approximation
where gravitational back reaction is neglected and, in particular, in which the
horizon is unchanged by the passage of our object.

We therefore first consider the thermal density matrix $\rho_{R0}$ which
results from tracing the Minkowski vacuum, $|0_M\rangle \langle 0_M|$,
over the invisible Rindler wedge:
\begin{equation}
\rho_{R0} = \mbox{Tr}_{invisible} |0_M \rangle \langle 0_M |.
\end{equation}
This describes all information that the Rindler observer can
access in the Minkowski vacuum state. We wish to compare $\rho_{R0}$ with another
density matrix $\rho_{R1}$ which provides the Rindler description of the
state $\rho_M$ above, in which one object (in an undetermined microstate) is added to
the Minkowski vacuum $|0_M \rangle$.  The density matrix $\rho_{R1}$ is
hence
\begin{equation}
\rho_{R1} = \mbox{Tr}_{invisible} \rho_M = \mbox{Tr}_{invisible}
\frac{1}{n} \sum_{i=1}^n |1_M;i \rangle \langle1_M;i|.
\end{equation}

We would like to compute the difference in energy
\begin{equation}
\delta E= \mbox{Tr} [H(\rho_{R1}-\rho_{R0})],
\end{equation}
and entropy
\begin{equation}
\delta S = - \mbox{Tr} [\rho_{R1} \ln \rho_{R1} - \rho_{R0} \ln \rho_{R0}],
\end{equation}
where $H$ is the Hamiltonian of the system and, in both cases, the sign has been
chosen so that the change is positive when $\rho_{R1}$ has the greater value of energy or
entropy.  The entropy of each state
separately is well known to be divergent \cite{entangle}.  However,
there is no reason to expect $\delta S$ to be ill-defined in our context.  
Assuming that the
object has some moderately well-defined inertially-measured 
energy $E_{inertial}$ and is well localized, the 
energy measured by the Rindler observer will also be reasonably well-defined
and the difference
$\rho_{R1} \ln \rho_{R1} - \rho_{R0} \ln \rho_{R0}$ will have negligibly small
diagonal entries at high energy so that the above trace will exist.
In other words, we may compute $\delta S$ by
first imposing a cutoff $\Lambda$, computing the entropy ($S_0,S_1$)
and energy ($E_0,E_1$) of the two states ($\rho_{R0},\rho_{R1}$) separately,
subtracting the results, and removing the cutoff.
We will see that the cutoff dependence is trivial in the approximation
used below.

Let us make the simplifying assumption that the object represents a
small perturbation on the initial thermal Rindler density matrix
$\rho_{R0}$; that is, $\rho_{R1} = \rho_{R0} + \delta \rho$, where ``$\delta
\rho \ll \rho_{R0}$''.  One expects this approximation to hold for $n e^{-E/T} \gg 1$, in
which case (on average) there are already many objects similar to ours present in
$\rho_{R0}$.  In particular, this approximation should hold in the limit of
large $n$ with fixed $E/T.$  A somewhat more careful argument for this approximation was explained to us by
Mark Srednicki and is given in the appendix.

We therefore approximate $\delta S$ by Taylor expanding
around $\rho_{R0}$,
\begin{equation}
\label{Sresult}
\delta S \approx - \mbox{Tr}\left[ \delta \rho {\delta (\rho \ln \rho)
    \over \delta \rho}\Big|_{\rho=\rho_{R0}} \right] = - \mbox{Tr} [
    \delta \rho (1 + \ln \rho_{R0})] = - \mbox{Tr} [\delta \rho \ln
    \rho_{R0}],
\end{equation}
where in the last step we use the fact that $\mbox{Tr} \rho_{R1} =
\mbox{Tr} \rho_{R0}$. Now recall that the initial density matrix is
thermal, $\rho_{R0} = e^{-H/T}/(\mbox{Tr} e^{-H/T})$; as a result 
\begin{equation}
\delta S \approx - \mbox{Tr} [ \delta \rho (-H/T)] = {\mbox{Tr} [H
    (\rho_{R1} - \rho_{R0})] \over T} = {\delta E \over T},
\end{equation}
where we have again used $\mbox{Tr}(\delta \rho)=0$.  This key result is independent of any cut-off.

This result can also be understood on the basis of classical thermodynamic
reasoning. The initial configuration $\rho_{R0}$ represents a thermal
equilibrium. We wish to calculate the change in entropy in a process
which increases the energy by an amount $\delta E$. Whatever the
nature of the object that we add, this cannot increase the entropy
by more than it would have increased had we added this energy as heat. Since
 we consider a small change in the configuration, the
first law yields
\begin{equation} \label{enbound}
\delta S < \delta S_{max} = {\delta E \over T}.
\end{equation}
We see that the process of adding a small object saturates this bound,
at least to first order in small quantities. We should expect that if
we relaxed the assumption that the object represents a small change in
the configuration, the resulting $\delta S$ will satisfy but no longer
saturate the integrated version of this bound.

We have found that the entropies ascribed to the localized object 
by Minkowski and Rindler
observers behave in quite different ways. From the point of view of
the Minkowski observer, entropy and energy are independent; the object
can have an arbitrarily large entropy with fixed energy if it has
sufficiently many internal states. On the other hand, from the point of view of the
Rindler observer,  entropy and energy are linked in
a very general way by the bound (\ref{enbound}).  As we will
see in the examples in the next section, the Rindler entropy for
`highly entropic objects' will be much smaller than the inertial
entropy. 

Note that since it arises from a trace over the invisible Rindler
wedge, this restriction is generally correlated with the presence of a
horizon, to which we might want to assign a geometric entropy. Thus,
this reduction in the ascribed entropy is important for a complete
understanding of attempts to violate the generalized second law using
such highly entropic objects, as reviewed in the introduction. We will
return to such issues in section \ref{disc}. 

\section{Calculations for free fields}

\label{calc}

We now explore two concrete examples to illustrate
and elucidate the general discussion above. We consider a system of
$n$ free bose or fermi fields, and calculate the entropy of objects
described, from the inertial point of view, by a single-particle density matrix uniformly
distributed over the different fields.  Thus the inertial observer assigns
the object an entropy $S_{inertial}= \ln n$ in each case.

\subsection{Bosons}

For a system of $n$ free bosonic fields, the Minkowski vacuum can be
recast in terms of the Rindler data as the following entangled state
\cite{unruh}:
\begin{equation}
|0_M\rangle = \prod_{i=1}^n \prod_j
(1-e^{-\frac{\omega_j}{T}})^{\frac{1}{2}} e^{-\frac{\omega_j}{2T}
  a_{ijL}^{\dagger}a_{ijR}^{\dagger} } |0_{Rindler}\rangle .
\end{equation}
Here $i$ labels the different fields, $j$ labels a complete set of modes
$u_{jL}, u_{jR}$ of positive Rindler frequency $\omega_j$ for each field, 
and the labels $R$ and $L$
refer respectively to the right and left Rindler wedges.  Each mode $u_{iL},u_{jR}$
of the $i$th field has an associated annihilation operator $a_{iL},a_{iR}$ which annihilates the Rindler
vacuum $|0_{Rindler}\rangle$ and satisfies a standard commutation relation of the form $a a^\dagger- a^\dagger a =1$.
The temperature $T$ associated with
the uniformly accelerating observer of interest is given by $T=\frac{a}{2\pi}
= \frac{\kappa}{2\pi}$ where 
$a$ is the observer's proper
acceleration and $\kappa$ is the surface gravity of the boost Killing field $\xi$
that is normalized on our observer's worldline. 

Similarly, the annihilation operator $a_{iM}$ 
for a Minkowski mode $u_M$ of the $i$th field can on general principles \cite{unruh}
be expressed in the form:
\begin{equation}
\label{Bog}
a_{iM} = \sum_{j}  \left[ (u_M,u_{jR})( a_{ijR} -
e^{-\frac{\omega_j}{2T}} a_{ijL}^{\dagger})
+ (u_M,u_{jL})( a_{ijL} -
e^{-\frac{\omega_j}{2T}} a_{ijR}^{\dagger})
\right],
\end{equation}
where $(u,v)$ is the Klein-Gordon inner product.  For simplicity we might suppose
that we choose a mode $u_M$ with no support on the invisible Rindler wedge (say, the left one) and for which
$(u_M,u_{jR})$ is well modeled by a delta-function; we will return to the general case later in the
subsection.  In particular, this simplification means that the Rindler frequency
$\omega$ of $u_M$ is reasonably well defined\footnote{\label{foot}If the object is well localized and located near our Rindler observer
at some time, then $\omega$ is also the frequency measured by the inertial observer.}.  For such a case the above Bogoliubov
transformation becomes
\begin{equation}
\label{trunc}
a_{iM} = \frac{1}{\sqrt{1-e^{-\frac{\omega}{T}}}}( a_{iR} -
e^{-\frac{\omega}{2T}} a_{iL}^{\dagger}),
\end{equation}
where the normalization of $a_{iM}$ is fixed by the commutation relation.
Here the Rindler operators refer to the one relevant pair of Rindler modes.

Since $a_{iM}$ acts in the Hilbert space describing only the $i$th field, it is
convenient to describe the action of $a_{iM}$ on the vacuum
$|0_{iM} \rangle$ for this particular field alone.  One sees that the
properly normalized Minkowski one-particle state is given
in terms of an infinite number of Rindler excitations
\begin{equation}
a_{iM}^{\dagger} |0_{iM}\rangle = (1-e^{-\frac{\omega}{T}}) \sum_k
e^{-\frac{k \omega}{2T}} \sqrt{k+1} |k, k+1\rangle_i,
\end{equation}
where $|k,k+1 \rangle_i$ denotes the state of field $i$ having $k$ Rindler
excitations in the right wedge and $k+1$ Rindler excitations in the left wedge.
Here we consider only the factor in the Hilbert space that describes
the modes appearing in (\ref{trunc}).

Thus, if we return to our Minkowski density matrix
\begin{equation}
\label{dm}
\rho_M = {1 \over n} \sum_{i=1}^n |1_M;i \rangle \langle1_M;i| = {1
  \over n} \sum_{i=1}^n (a_{iM}^{\dagger} |0_M\rangle \langle0_M|
  a_{iM}), 
\end{equation}
tracing over the invisible Rindler wedge will give the desired result.
A short calculation gives 
\begin{equation}
\rho^{diag}_{R1} = \mbox{Tr}_{invisible} \rho_M = (e^{\frac{\omega}{T}}-1) \frac{1}{n}\sum_i N_i \rho_{R0},
\end{equation}
where $\rho_{R0}$ is the original density matrix corresponding to the
Minkowski vacuum, $\rho_{R0} = \mbox{Tr}_{invisible} |0_M\rangle
\langle0_M|$, and $N_i$ denotes the number operator for the $i$th field. 
The superscript $diag$ indicates that we have written only the diagonal part of $\rho_{R1}$ in the
standard basis, which the reader will shortly see is all that contributes to $\delta E$ and $\delta S$ below.  One can check
explicitly that $\rho_{R1}$ is properly normalized so that its trace is
1. 

The change in the density matrix is thus $\delta \rho = \rho_{R1} -
\rho_{R0}$.  As in the previous section, we assume that this
change is small, so that we can compute the change in entropy by $\delta S = -
Tr (\delta \rho (1+ \ln \rho))$. For non-interacting particles, the
Hamiltonian is $H = \sum_i N_i \omega$, so the average change in energy is
\begin{equation}
\label{dE}
\delta E = \omega \mbox{Tr} (\sum_i N_i \delta \rho) =
\frac {\omega}{1-  e^{-\frac{\omega}{T}}}. 
\end{equation}

This result is already of interest.  Note that for $\omega \gg T$ and
under the conditions of footnote \ref{foot}, one finds $\delta E =
E_{inertial} \approx \omega$.  On the other hand, for $\omega \ll T$
the background of objects in the thermal bath leads to an
amplification reminiscent of the effect of stimulated
emission\footnote{Interesting comments on the emergence of the Rindler
  thermal fluctuations can be found in \cite{sciama}.}.

On the other hand, the change in entropy is 
\begin{equation}
\label{dS}
\delta S = - \mbox{Tr}[\delta \rho (1 +\ln \rho_{R0})] = - \mbox{Tr} [\delta \rho (1 +C
- \frac{\omega N}{T})] = \frac{\delta E}{T},
\end{equation}
where in the second step we used that $\rho_{R0}$ is thermal; (i.e.,  $\rho_{R0} = C
e^{-H \over T}$, where $C$ is a c-number), and in the final step we use
$\mbox{Tr} \delta \rho =0$.   It is clear that (\ref{dE}) and (\ref{dS}) depend only on the diagonal part of $\rho_{R1}$
since the other factors in the trace are all diagonal.

This expression obviously saturates the
bound provided by the second law of thermodynamics in agreement with the general arguments of
section \ref{gen}.
The results are independent of the number of species $n$ due to the fact that both 
$\delta E$ and $\delta S$ are linear in $\delta \rho$, while $\delta
\rho$ is a normalized average
over terms in which each field is changed independently.  Thus for large $n$ the Rindler observer will ascribe a much smaller 
entropy to the object.

Finally, let us return to the case where more than one pair of Rindler modes contributes to (\ref{Bog}).
Since $\rho_M$ is quadratic in $a_{iM}$, this leads to two sums over Rindler operators in (\ref{dm}).
However, since $N_i$ and $(1 + \ln \rho_{R0})$ are diagonal in the standard Rindler Fock basis, 
inspection of (\ref{dE}) and (\ref{dS}) shows that only the corresponding diagonal part of $\rho_M$ will 
contribute to $\delta E$ and $\delta S$.  This diagonal part contains only a single sum over modes, weighted
by $|(u_M,u_{jR})|^2$ or $|(u_M,u_{jL})|^2$.  In the case where the Minkowski mode has no support in the left
wedge (so that  $|(u_M,u_{jL})|^2$ =0), the effect is just as if each mode represented a separate field.
But we have seen that the final result is independent of the number of fields and, by the same logic, it is independent
of the weighting given to to the various fields in (\ref{dm}), so long as $\rho_M$ is properly normalized.
Thus we again obtain (\ref{dE}) and (\ref{dS}).  In fact, it is clear that spreading the support over many modes
only helps to justify our approximation as it increases the effective number of fields.  In the more general case where
the Minkowski mode function overlaps the invisible wedge, the quantities $\delta E$ and $\delta S$ will be reduced
in proportion to the probability that the object lies in the invisible wedge but one again finds $\delta S = \delta E/T$.

\subsection{Fermions}

For non-interacting fermions the calculation proceeds in a
very similar fashion.  The Minkowski vacuum written as an entangled
state is now
\begin{equation}
|0_M\rangle = \prod_i \prod_j (1+e^{-\frac{\omega_j}{T}})^{\frac{1}{2}} 
e^{-\frac{\omega_j}{2T} b_{ijL}^{\dagger}b_{ijR}^{\dagger} } |0_{Rindler}\rangle, 
\end{equation}
where the only changes from before are that the $b$ operators are
fermionic with $bb^\dagger + b^\dagger b = 1$, and the associated change in
sign in the thermal normalization factor.  

As for the boson field, working to first order in $\delta \rho$ makes $\delta E$ 
and $\delta S$ independent of the number of species $n$.  As a result, we may take $n=1$ to evaluate the linearized result.
Though the
linearized approximation is in general valid only for large $n$, this does not prevent us from calculating the linearized
result for $n=1$ and using it to correctly compute the large $n$
result. Similarly, since only the diagonal part of $\delta \rho$ will contribute we will also obtain the correct result (for an object whose wavefunction vanishes in the invisible wedge) by supposing that
our Minkowski mode overlaps only a single pair of Rindler modes. 
The associated Bogoliubov
transformation then takes the form
\begin{equation}
b_M = \frac{1}{\sqrt{1+e^{-\frac{\omega}{T}}}}( b_L + e^{-\frac{\omega}{2T}} 
b_R^{\dagger}).
\end{equation}
For a single field, the Rindler description of the Minkowskian one particle state is simply
\begin{equation}
b_M^{\dagger} |0_M\rangle = |1,0\rangle_{Rindler},
\end{equation}
with one particle in the visible wedge and none in the invisible wedge.
The properly normalized density matrix is just $\rho_{R1} =
|1\rangle\langle1|$.  Hence, the change in the Rindler density matrix
$\delta\rho$ is
\begin{equation}
\delta \rho = -(1+e^{-\frac{\omega}{T}})^{-1}(|0\rangle \langle 0| -
|1 \rangle\langle 1|),
\end{equation}
and the change in energy for a non-interacting system of excitations is
\begin{equation}
\mbox{Tr}(H\delta \rho) = \frac{\omega}{1+e^{-\frac{\omega}{T}}}.
\end{equation}
In general, one now finds $\delta E < E_{inertial}$ so that $\delta S \ll S_{inertial}$ for large $n$.
When the assumptions of footnote \ref{foot} hold and $\omega \gg T$, 
one also finds $\delta E = E_{inertial} \approx \omega$, but for $\omega \ll T$, the background thermal bath of Fermions leads to
a {\it suppression} of the Rindler energy.  Said differently, the appearance of an object from the inertial
point of view can sometimes correspond to the disappearance of an object from the Rindler observer's
thermal bath.

Since $\delta \rho$ is traceless, one again finds $\delta S = \frac{\delta E}{T}$, 
saturating the bound provided by the first and second laws.  

\section{Discussion}

\label{disc}

We have argued that inertial and accelerated observers naturally ascribe different values to the entropy of a localized
matter system in flat Minkowski space, even if the system is fully visible to both observers and the observers describe the
system with the same resolution.  Certainly, the observers measure a different $\delta S$ when the localized system is
introduced or removed, and so must use different values when considering a thermodynamic accounting of such processes.
The results support the suggestions of \cite{Wald1} that entropy, {\it even of ordinary matter systems}, is a very
subtle concept in general relativity or quantum field theory in curved spacetime.  

What is surprising from the point of view
of \cite{Wald1} is that our effect occurs even when the observers agree on the {\it energy} of each microstate.
Our effect is in some sense due to a ``mixing" between the object considered
and the thermal background seen by the Rindler observer.
In particular, we argued that when the number of microstates $n$ satisfies $n e^{-E/T} \gg 1$ the entropy assigned to our object by the
accelerated observer is $\delta S = \delta E/T$.  Note that in this regime a thermal ensemble of {\it distinguishable} particles would
diverge, as each new particle would add an entropy $\ln n \gg E/T$.  Thus, statistics plays an important role and one cannot
rely on any intuition built from the study of distinguishable particles.

The above sections provided general arguments as well as explicit calculations for free fields.
It is clear that the general arguments apply equally well to static observers in the presence of a black hole
or a cosmological (e.g., de Sitter) horizon, and that the free field calculations would take a form that is essentially identical.
Our general arguments are quite similar to those of \cite{sorkin}, and 
our observer dependence of entropy provides an important physical mechanism behind such
arguments.
We note that the theorem derived in \cite{sorkin} applies
rigorously in the context considered here in which the temperature and the location of the horizon are held constant and the
dependence on any cutoff is trivial.

In the black hole case, the result $\delta S = \delta E/T$ would apply to any static observer and
would in general differ from the entropy assigned by a freely falling observer.  
It is therefore interesting to return to the discussion of attempts to violate the generalized second law (i.e., including horizon entropy)
by dropping an object with $n \gg e^{E/T}$ microstates into a black hole.  Suppose first that the object
begins far away and that we work in asymptotically flat space.  Then the object will begin in a region of space that is
{\it not} in thermal equilibrium with the black hole and where the effective local temperature vanishes.
Here a static observer measures the object to have entropy $\ln n$ since there is no thermal bath.
On the other hand, as the object approaches the black hole horizon one expects that it enters a region which is
(nearly) at thermal equilibrium at the black hole temperature $T$ as described by the distant static observer.
If the object acts like a free field\footnote{And thus falls freely so that the energy is constant in the test object
approximation.}, we have seen that at this point the static observer can ascribe only
an entropy $S = E/T$ to the object.  Thus {\it if nothing else occurs} the second law will indeed have been violated.
The important point, however, is that this will occur {\it before} the object reaches the horizon itself.
Thus, one expects that the generalized second law is protected by the same mechanism that protects the ordinary
second law.  In the present case, one expects to see a large flux of objects radiated by the black hole as, inverting the above
argument, movement of highly entropic objects from the warm near-horizon neighborhood into the cold region of space will
dramatically increase the entropy!  

The effect is similar to attempting to send a low frequency photon (with two internal states)
into a hot cavity which is otherwise at equilibrium -- many more photons will emerge.  Sending in a larger
flux of photons does not help as a beam with $N$ photons does not in general have entropy $N \ln 2$ unless the photons
are well separated, but well-separated photons will never overpower the large outward flux.  This latter effect
is also important for AdS black holes which can be at equilibrium with the entire spacetime.

As with \cite{sorkin}, this result suggests that the generalized second law holds whenever the region outside
the black hole is described by standard quantum field theory, regardless of whether one imposes bounds of the forms
suggested in \cite{Bek,erice,LS,tHooft}.  
However, both stop short of being proofs because they do
not trace the actual dynamical changes as an object falls through the horizon.  The absorption of the object by the
black hole will change both the temperature and the size of the horizon.  Since a change of the horizon size in some sense
implies a change in the ``location" of the horizon, any proof built without a detailed theory of quantum gravity would
likely require careful assumptions as to the treatment of modes near the horizon which, by themselves, lead
to a divergence in the total entropy $-\mbox{Tr} \rho \ln \rho$ of the thermal bath seen by the accelerated observer \cite{entangle}.
For a finite (as opposed to infinitesimal) process, careful consideration would need to be given to non-equilibrium issues.
Some analysis of the effect on the horizon was reported in \cite{JP}, but further progress in this direction would be of use.

{\bf Acknowledgments:}
The stimulating atmosphere of the Aspen Center of Physics,
where this work was initiated, is gratefully acknowledged.  D.M. would also like to thank Ted Jacobson and Mark Srednicki for 
interesting discussions.
D.M. was supported in part by NSF grant PHY03-54978, and by funds from
the University of California. S.F.R. was supported by the EPSRC.

\appendix

\section{Evaluation of $\delta S$ for large $n$}

Since $\delta E$ is exactly linear in $\delta \rho$, justification of our approximation requires only showing that $\delta S$ takes the form 
(\ref{Sresult}) for large $n$.  This can be argued through a standard trick from statistical mechanics\footnote{We
thank Mark Srednicki for explaining this argument to us.}.  One notes that the entropy $S = - \mbox{Tr} (\rho \ln \rho)$ is
$-1$ times the order $\epsilon$ term in the expansion of $\mbox{Tr} \rho^{1+ \epsilon} = \mbox{Tr} \rho e^{\epsilon \ln \rho}$.

Let us now compute the entropy of $\rho_{R1}$.  We note that this
density matrix is an average of $n$ terms, each 
identical to $\rho_{R0}$ except that the density matrix describing a certain subsystem has been changed.
The average is over the label telling us {\it which} subsystem has been changed.  Thus, if $\rho_{R0}$
takes the form $\rho_{R0} = \otimes_{i=1}^n A$, then we may write

\begin{equation}
\rho_{R1} =  \frac{1}{n} \sum_i A \otimes ... \otimes A \otimes B \otimes A \otimes ... \otimes A
\end{equation} where again the index $i$
tells us where to place the factor $B$.

It is straightforward to see that for integer $k \ll n$ we have 

\begin{equation}
\mbox{Tr} \rho_{R1} =  \left[ \mbox{Tr}(A^{k-1}B) \right]^k  \left[ \mbox{Tr}(A^{k}) \right]^{n-k} + O(1/n),
\end{equation} 
since the number of terms having more than one factor of $B$ in the same subsystem are $O(1/n)$.
The standard trick is to substitute $k = 1+ \epsilon$ and take the coefficient of $\epsilon$ as ($-1$ times) the entropy.
This yields
\begin{equation}
\mbox{Tr} \rho_{R1} =  \left[ \mbox{Tr}[(1 + \epsilon \ln A )B] \right]^{1 + \epsilon}  \left[ \mbox{Tr}(A(1 + \epsilon \ln A)) \right]^{n-1 - \epsilon}.
\end{equation}
But $\mbox{Tr} A = \mbox{Tr} B = 1$.  So we have
\begin{equation}
\mbox{Tr} \rho_{R1} =  \left[ 1 + \epsilon  \mbox{Tr} (B \ln A ) \right]   \left[ 1 + \epsilon (n-1) \mbox{Tr} (A\ln A) \right]
\approx 1 + \epsilon \left[ \mbox{Tr} [(B-A) \ln A] + n \mbox{Tr} (A \ln A) \right].
\end{equation}
Recognizing $n \mbox{Tr} (A \ln A)$ as $-1$ times the entropy of $\rho_0$, it follows that
\begin{equation}
\delta S = - \mbox{Tr} [(B-A) \ln A] = - \mbox{Tr} \delta \rho \ln \rho_{R0},
\end{equation}
where in the last step we have again used $\mbox{Tr}(A-B)=0$.

\end{document}